\documentclass[%
aip,jap,
 amsmath,amssymb,floatfix,
 preprint,%
]{revtex4-1}
\usepackage{multirow}
\usepackage{amsmath,amssymb,amsfonts}
\usepackage{algorithmic}
\usepackage{array}
\usepackage{xcolor}
\usepackage{footnote}
\usepackage{threeparttable}
\usepackage{graphicx}
\usepackage{textcomp}
\usepackage{xcolor}
\usepackage{array}
\usepackage{mathtools}
\usepackage{siunitx}

\DeclarePairedDelimiter\abs{\lvert}{\rvert}%

\def\BibTeX{{\rm B\kern-.05em{\sc i\kern-.025em b}\kern-.08em
    T\kern-.1667em\lower.7ex\hbox{E}\kern-.125emX}}
\begin{document}

\title{Spin Wave Based Full Adder}
\author{Abdulqader Mahmoud}
\affiliation{Delft University of Technology, Department of Quantum and Computer Engineering, 2628 CD Delft, The Netherlands}
\email{a.n.n.mahmoud@tudelft.nl}

\author{Frederic Vanderveken}
\affiliation{KU Leuven, Department of Materials, SIEM, 3001 Leuven, Belgium}
\affiliation{Imec, 3001 Leuven, Belgium}
\author{Florin Ciubotaru}
\affiliation{Imec, 3001 Leuven, Belgium}

\author{Christoph Adelmann}
\affiliation{Imec, 3001 Leuven, Belgium}

\author{Sorin Cotofana}
\affiliation{Delft University of Technology, Department of Quantum and Computer Engineering, 2628 CD Delft, The Netherlands}

\author{Said Hamdioui}
\affiliation{Delft University of Technology, Department of Quantum and Computer Engineering, 2628 CD Delft, The Netherlands}

\begin{abstract}
Spin Waves (SWs) propagate through  magnetic waveguides and interfere with each other without consuming noticeable energy, which opens the road to new ultra-low energy circuit designs. 
In this paper we build upon SW features and propose a novel energy efficient Full Adder (FA) design consisting of $1$ Majority  and $2$ XOR gates, which outputs $Sum$ and $Carry-out$ are generated by means of threshold and phase detection, respectively. We validate our proposal by means of MuMax3 micromagnetic simulations and we evaluate and compare its performance with state-of-the-art SW, \SI{22}{nm} CMOS, Magnetic Tunnel Junction (MTJ), Spin Hall Effect (SHE), Domain Wall Motion (DWM), and Spin-CMOS implementations. Our evaluation indicates that the proposed SW FA consumes $22.5$\% and $43$\% less energy than the direct SW gate based and \SI{22}{nm} CMOS counterparts, respectively. Moreover it exhibits a more than $3$ orders of magnitude smaller energy consumption when compared with state-of-the-art MTJ, SHE, DWM, and Spin-CMOS based FAs, and outperforms  its contenders in terms of area by requiring at least $22$\% less chip real-estate.

\end{abstract}

\maketitle

\section{Introduction}

The raw data amount has increased rapidly in the last $20$ years because of the information technology revolution and its need for highly efficient computing platforms \cite{data1}. To satisfy these requirements, CMOS has been strongly downscaled to further improve its performance \cite{ITRS}. However, because of three main walls \cite{cmosscaling2}: (i) leakage wall, (ii) reliability wall \cite{cmosscaling1}, and (iii) cost wall, it becomes very difficult to further downscale CMOS, which indicates the near (economical) end of Moore's law. Therefore, multiple other technologies have been explored, e.g., memristors \cite{memristor10}, and spintronics \cite{ITRS} with the hope to further improve computer performance. One of the most efficient spintronics technologies is the Spin Wave (SW) because of \cite{SW,SW1}: (i) its ultra-low energy consumption as the charge doesn't move; (ii) its acceptable delay; and (iii) its wavelength can reach the nanometer scale. Hence, designing spin wave circuits, e.g., FAs, is of great interest to enable building spin wave computers.

Research on SW technology based logic and circuit designs is in early stage. At the logic/gate level, some basic single output gates (such as NOT, (N)AND, (N)OR, and X(N)OR) were reported in \cite{logic21,logic11,logic12}, while some multi-output gates were discussed in \cite{fanout,fanout10,SW2}. At the circuit level, dedicated operators for neuromorphic applications were developed in \cite{nonboolean1,neuromorphic_computing}; examples are upper and lower threshold operators, truncated difference operators, literal operators, cyclic operators and minimum and maximum operators. In addition, exploring the concept of wave pipelining based operation was illustrated in \cite{Pipeline}. Further, design for arithmetic operation such as FA was explained in \cite{logic1}. However, the designs in \cite{nonboolean1,neuromorphic_computing,Pipeline,logic1} were reported at the conceptual level without any validation. Preliminary demonstrators were presented in \cite{MUX,memory3,memory5}; these include $\mu$m range $2$ to $1$ mutliplexer and $mm$ range Magnonic Helographic Memory (MHM), respectively. In conclusion, clearly circuit designs for spin wave computing is in its infancy stage; efficient designs at different scales of complexity should still be developed, validated and demonstrated in order to set up a step towards spin wave computing engines.

This paper proposes and validates a novel SW FA. The adder is based on two SW gates where the outputs are generated using two different mechanisms; threshold detection and phase detection. This work main contributions can be summarized as follows: 
\begin{itemize}

  \item Developing and designing a SW FA: a Majority gate and $2$ XOR gates are utilized to build the FA based while threshold and phase detection are utilized to capture the $Sum$ and $Carry-out$ outputs, respectively. 

  \item Validating the proposed FA: MuMax3 software is utilized to validate the correct behavior of the proposed FA. 
  
   \item Demonstrating the superiority: we assess the proposed FA and compare it with  state-of-the-art SW, \SI{22}{nm} CMOS, Magnetic Tunnel Junction (MTJ), Spin Hall Effect (SHE), Domain Wall Motion (DWM), and Spin-CMOS implementations. Our evaluation indicates that the proposed SW FA consumes $22.5$\% and $43$\% less energy than the direct SW gate based and \SI{22}{nm} CMOS counterparts, respectively. Moreover it exhibits a more than $3$ orders of magnitude smaller energy consumption when compared with state-of-the-art MTJ, SHE, DWM, and Spin-CMOS based FAs, and outperforms  its contenders in terms of area by requiring at least $22$\% less chip real-estate.  
\end{itemize}

The rest of the paper is organized as follows. Section \ref{sec:Basics of spin-wave technology} explains the SW fundamentals and SW computing paradigm. Section \ref{sec:Proposed Spin Wave Based FA} illustrates the proposed SW FA. Section \ref{sec:Simulation Setup and Results} gives the simulation setup, and the performed simulation. Section \ref{sec:Discussion} estimates the energy consumption of the proposed FA, compares it with the state-of-the-art counterparts, and provides some inside on the impact of variability and thermal noise effects. Section \ref{sec:Conclusion} concludes the paper. 

\section{SW technology background}
\label{sec:Basics of spin-wave technology}

Spintronic devices, such as spin wave based, exploit the magnetization state and its dynamic behavior to implement their functionality. This magnetization dynmaics can be described by the Landau-Lifshitz-Gilbert (LLG) Equation \cite{LL_eq}\cite{G_eq}:$\frac{d\vec{M}}{dt} =-\abs{\gamma} \mu_0 \left (\vec{M} \times \vec{H}_{eff} \right ) + \frac{\alpha}{M_s} \left (\vec{M} \times \frac{d\vec{M}}{dt}\right ),$ where $\gamma$ is the gyromagnetic ratio, $\mu_0$ the vacuum permeability, $\alpha$ the Gilbert damping constant, $\vec{M}$ the magnetization, $M_s$ the saturation magnetization, and $\vec{H}_{eff}$ the effective field. In this work, we consider the effective field as the sum of the external field, exchange field, demagnetization field and magnetocrystalline anisotropy field. 

A weak perturbation of the magnetization equilibrium state can be described by the linearised LLG equation. This linearised LLG equation has wave-like solutions which are known as SWs. These solutions span over the full magnetic volume, and therefore, SWs are also defined as collective magnetization excitations in the magnetic materials \cite{SW}. 

The spin wave computing paradigm is based on the wave interference principle, which enables the direct implementation of logic gates without the need for the traditional Boolean algebra formalism \cite{SW}. In a general way, if multiple spin waves coexist in the same waveguide, they interfere with each other depending on their amplitude, wavelength, phase, and frequency \cite{SW,parallelism}. For example, the interference of two spin waves that have the same amplitude, wavelength, and frequency is considered. If these two spin waves have the same phase, then they interfere constructively resulting in a wave with larger amplitude. When the two waves have opposite phases, then they interfere destructively and cancel each other resulting in a zero amplitude \cite{SW}. 
In addition, if an odd number of SWs interfere, the interference result is based on the majority principle. For example, if $3$ SWs with the same amplitude, wavelength, and frequency coexist in the same waveguide, then the resultant spin wave has a phase of $0$ if at least $2$ SWs have phase of $0$, whereas the resultant spin wave has a phase of $\pi$ if at least $2$ SWs have phase of $\pi$ \cite{SW}. Furthermore, we note that a $3$-input Majority gate implementation requires $18$ transistors in CMOS, whereas it is implemented in the SW domain by the interference of $3$ SWs in a single waveguide \cite{SW}.

\section{Proposed Spin Wave Based Full Adder}
\label{sec:Proposed Spin Wave Based FA}

\begin{figure}[t]
\centering
  \includegraphics[width=0.7\linewidth]{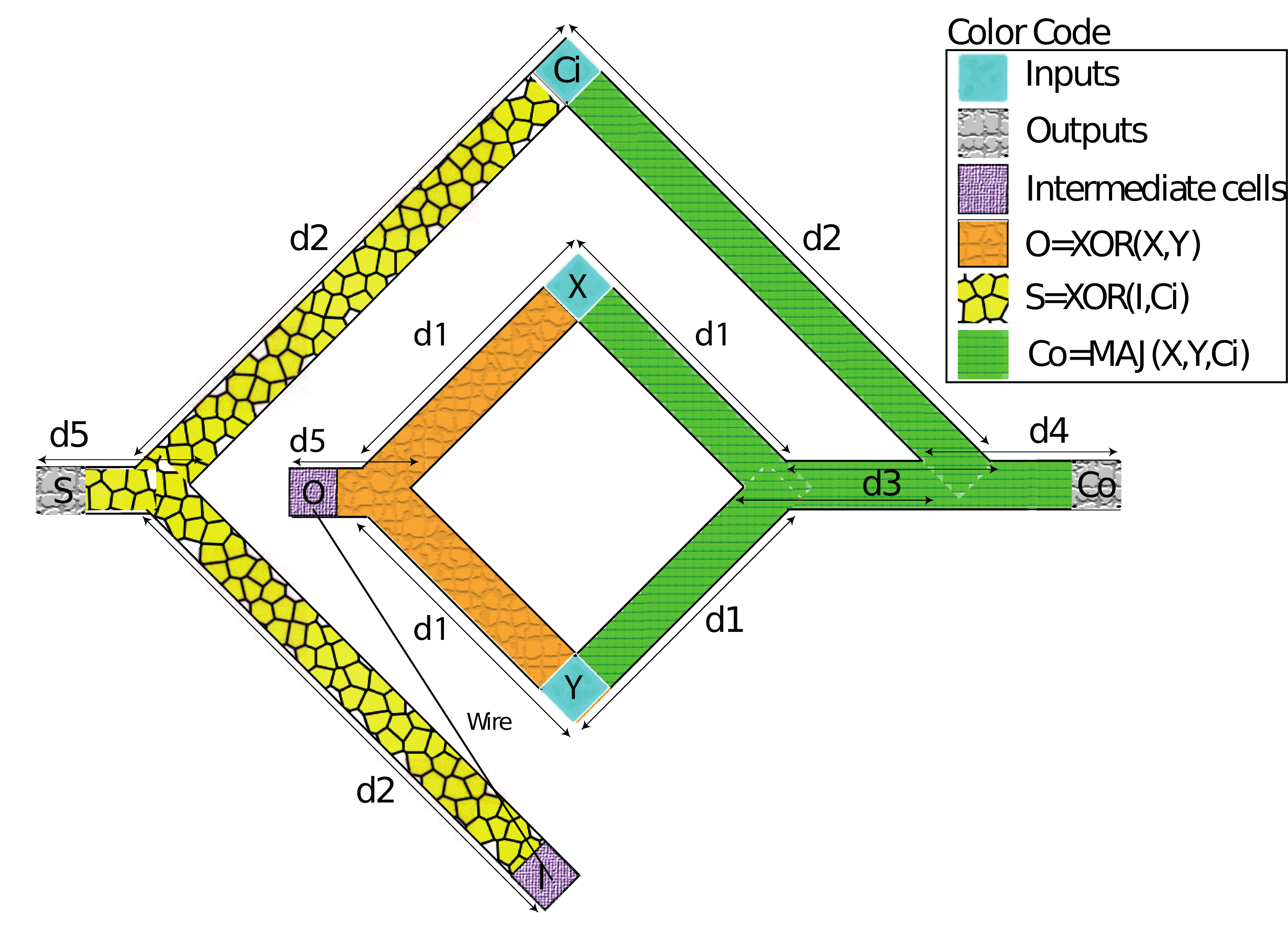}
  \caption{Spin Wave Based Full Adder.}
  \label{fig:structure1}
\end{figure}

Figure \ref{fig:structure1} presents the novel developed energy efficient $1$-bit FA structure with inputs $X$, $Y$, and carry-in $C_i$, and outputs Sum $S$ and Carry-out $C_o$. It is implemented by utilizing two XOR gates, and one Majority gate. The XOR gates are used to determine the Sum output and the Majority gate is used to determine the Carry-out output. The output of the first XOR gate being $O=XOR(X,Y)$ is fed into the second XOR together with $C_i$ to produce the FA Sum $S=XOR(I,C_i)$. Note that $O$ is connected to $I$ by a metal wire that allows the excitation of a spin waves at $I$ with the same phase as the one detected at $O$. That Majority gate is used to generate carry-out $C_o=MAJ(X,Y,C_i)$. The FA's excitation and detection cells can be voltage driven or current driven cells depending on the utilized excitation and detection methods. Different options for the spin wave excitation and detection can be used such as magnetoelectric cells \cite{SW,excitation1}, microstrip antennas \cite{SW,ref101},  and spin orbit torques \cite{SW,ref100}.

The FA parameters must be carefully designed in order to achieve the desired functionality. The waveguide width must be less than the SW wavelength $\lambda$ in order to have a proper interference pattern. In addition, all SWs must be excited with the \textit{same} amplitude, wavelength, and frequency to guarantee the desired SWs interference results. Moreover, the waveguide's length must be chosen accurately to obtain the desired outputs. For example, if SWs with the same phase have to interfere \textit{constructively} and SWs with opposite phase have to interfere \textit{destructively}, then the distances $d_1$ and $d_2$ must be equal to $n\lambda$ (where $n = 0,1,2,3,\ldots$). In the other case, when SWs with the same phase have to interfere \textit{destructively} and SWs with opposite phase have to interfere \textit{constructively}, then the distances $d_1$ and $d_2$ must be equal to $(n+1/2) \lambda$. 

Two main techniques are available to detect the spin wave output, namely phase detection and threshold detection. Phase detection detects the phase of the spin wave and compares it with a  predefined value. If the phase difference between the detected and the predefined phase is $0$, then the output is logic $0$, whereas if the phase difference is $\pi$, then the output is logic $1$. On the other hand, threshold detection detects the spin wave amplitude and compares it with a predefined value. If the spin wave amplitude is larger than the predefined threshold, then the output is logic $0$, whereas the output is logic $1$ if the spin wave amplitude is less than or equal to the predefined threshold. When phase detection is used, the distances $d_4$ and $d_5$ must be chosen accurately because both the non-inverted and the inverted versions can be detected depending on the distance between the output and the last interference point. For instance, if the desired result is to capture the non-inverted output, $d_4$ and $d_5$ must be $n\lambda$, whereas $d_4$ and $d_5$ must be $(n+1/2) \lambda$ if the inverted output is desired. On the other hand, if the threshold detection is utilized, the distances $d_4$ and $d_5$ must be as close as possible to the last interference point in order to detect large spin wave amplitude as this is crucial during the threshold detection.

To detect the outputs $S$ and $C_o$  (see Figure \ref{fig:structure1}) correctly the proposed FA operates as follows: 
\begin{itemize}
 \item Sum $S$: SWs excited at $X$ and $Y$ interfere with each other and the resultant SW is detected at $O$ based on threshold detection. Next, the detected output at $O$ feeds the input of the second XOR gate by exciting a SW with suitable phase. Finally, the SWs excited at $I$ and $C_i$ interfere with each other and the resultant SW is detected at $S$ based on the threshold detection. 
 \item Carry out $C_o$: The excited SWs at $X$ and $Y$ interfere constructively or destructively with each other depending on their phases. Then the resultant SW propagates and interferes with the excited SW at $C_i$. Finally, the phase of the resulting SW is detected at $C_o$.
 \end{itemize}
\section{Simulation Setup and Results}
\label{sec:Simulation Setup and Results}
In this section, we explain the simulation setup, the performed experiments, and their  results.
\subsection{Simulation Setup}
We made use of $w=\SI{50}{nm}$ wide  $Fe_{60}Co_{20}B_{20}$ waveguide to validate the proposed FA by means of MuMax3 \cite{mumax} with the parameters specified in Table \ref{table:1} \cite{parameters}. There is no need for an out-of-plane external field as the perpendicular magnetic anisotropy cants the device magnetization in the out-of-plane direction. We set up the SW wavelength $\lambda$ to be \SI{55}{nm}, which is larger than the waveguide width. Based on this, optimal design device dimensions are calculated resulting into: $d_1$=\SI{330}{nm} ($n=6$), $d_2$=\SI{880}{nm} ($n=16$), $d_3$=\SI{220}{nm} ($n=4$), $d_4$=\SI{55}{nm} ($n=1$), and $d_5$=\SI{55}{nm} ($n=1$). To calculate the SW frequency,  first the SW dispersion relation \cite{dispersionrelation} is determined; this is done based on the parameters of Table I and the waveguide width. 
From the FVSW dispersion relation and by setting the wavenumber to be $k$=$2\pi/\lambda$=\SI{50}{rad/\mu m}, the frequency is derived to be  $f=\SI{10}{GHz}$.

\begin{table}[t]
\caption{Parameters.}
\label{table:1}
\centering
  \begin{tabular}{|c|c|}
    \hline
    Parameters & Values \\
    \hline
    Magnetic saturation $M_s$ & $1.1$ $\times$ $10^6$ A/m \\
    \hline
    Perpendicular anisotropy constant $k_{ani}$ & $8.3$ $\times$ $10^5$ J/$m^3$\\
    \hline
    damping constant $\alpha$ & $0.004$ \\
    \hline
    Exchange stiffness $A_{exch}$ & $18.5$ pJ/m \\
    \hline
    Thickness $t$ & \SI{1}{nm} \\
    \hline
  \end{tabular}
\end{table}

\subsection{Performed Simulation}

Table \ref{table:2} presents the normalized magnetization values of the FA's Sum $S$ output for different input combinations \{$X$,$Y$,$C_i$\}= \{$0$,$0$,$0$\}, \{$0$,$0$,$0$\}, \{$0$,$0$,$1$\}, \{$0$,$1$,$0$\}, \{$0$,$1$,$1$\}, \{$1$,$0$,$0$\}, \{$1$,$0$,$1$\}, \{$1$,$1$,$0$\}, and \{$1$,$1$,$1$\}, respectively. Note that threshold detection is used to generate the output $S$. As can be observed from the Table, the first intermediate cell $O$, which is the XOR of $X$ and $Y$, can be implemented by choosing a suitable threshold such that if $O$ is greater than the threshold $O=0$, whereas $O=1$ otherwise. The appropriate threshold in this case is $0.515$, which is the average of $1$ and $0.03$. In this case, $O=0$ for the inputs combinations \{$X$,$Y$\}= \{$0$,$0$\} and \{$1$,$1$\}, whereas $O=1$ for the inputs combinations \{$X$,$Y$\}=\{$0$,$1$\} and \{$1$,$0$\}. As mentioned previously, the phase of the second intermediate cell $I$ equals to the phase of the first intermediate cell $O$. To generate the output $S$, which is realized by the XOR of $I$ and $C_i$, a new threshold should be selected; this should be the average of $0.98$ and $0.59$, resulting in a threshold of $0.785$. In this case, $S=0$ for the inputs combinations \{$I$,$C_i$\}=\{$0$,$0$\} and \{$1$,$1$\}, whereas $S=1$ for the inputs combinations \{$I$,$C_i$\}=\{$0$,$1$\} and \{$1$,$0$\}, which reflects the correct detection of the FA Sum output. Hence, the simulation validates the correct generation of the Sum output of the FA when appropriate thresholds are selected. 

\begin{table}[t]
\caption{Normalized Full Adder Sum Output Magnetization.}
\label{table:2}
\centering
  \begin{tabular}{|c|c|c|c|c|c|}
    \hline
    $C_i$ & $X$ & $Y$ & $O$  & $I$  & $S$ \\ 
    \hline
    $0$ & $0$ & $0$ & $1$ & $0$ & $0.98$ \\
    \hline
    $0$ & $0$ & $1$ & $0.03$ & $1$ & $0.59$ \\
    \hline
    $0$ & $1$ & $0$ & $0.03$ & $1$ & $0.58$ \\
    \hline
    $0$ & $1$ & $1$ & $1$ & $0$ & $1$ \\
    \hline
    $1$ & $0$ & $0$ & $1$ & $0$ & $0.59$ \\
    \hline
    $1$ & $0$ & $1$ & $0.03$ & $1$ & $1$ \\
    \hline
    $1$ & $1$ & $0$ & $0.03$ & $1$ & $0.99$ \\
    \hline
    $1$ & $1$ & $1$ & $1$ & $0$ & $0.58$ \\
    \hline
  \end{tabular}
\end{table}

\begin{figure}[t]
\centering
  \includegraphics[width=0.8\linewidth]{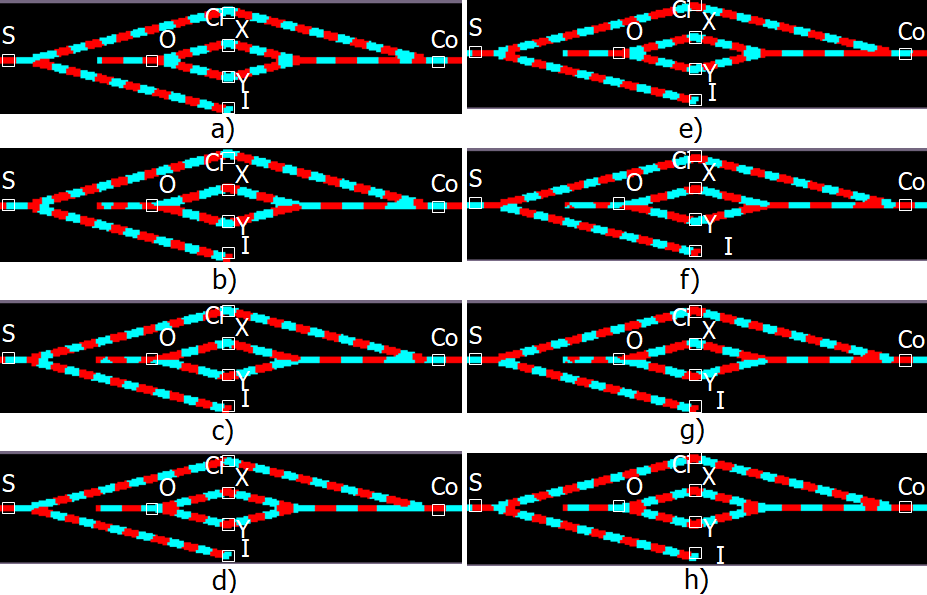}
  \caption{Spin Wave Based FA MuMax3 Simulation.}
  \label{fig:results1}
\end{figure}

Figure \ref{fig:results1} a) to h) present the results of the proposed FA Carry-out $C_o$ output for different input combinations \{$X$,$Y$,$C_i$\}= \{$0$,$0$,$0$\}, \{$0$,$0$,$0$\}, \{$0$,$0$,$1$\}, \{$0$,$1$,$0$\}, \{$0$,$1$,$1$\}, \{$1$,$0$,$0$\}, \{$1$,$0$,$1$\}, \{$1$,$1$,$0$\}, and \{$1$,$1$,$1$\}, respectively. In the Figure, the blue color represents logic $0$ whereas the red color represents logic $1$ and indicates that the output $C_{o}$ of the adder is correctly captured. For instance, $C_o=0$ for the input combinations \{$X$,$Y$,$C_i$\}= \{$0$,$0$,$0$\}, \{$0$,$0$,$1$\}, \{$0$,$1$,$0$\}, and \{$1$,$0$,$0$\}, whereas $C_o=1$ for the input patterns \{$X$,$Y$,$C_i$\}= \{$0$,$1$,$1$\}, \{$1$,$0$,$1$\}, \{$1$,$1$,$0$\}, and \{$1$,$1$,$1$\}, which proves that the FA Carry-out output is correctly generated. Note that although Sum output is presented in the Figure its colour is not relevant as threshold based detection is in place for it (see Table \ref{table:2}).

In conclusion, the simulation results demonstrate that by combining threshold detection and phase detection, a $1$-bit FA can be designed.

\section{Performance Evaluation and Discussion}
\label{sec:Discussion}

In this section we assess and compare the proposed FA and a number of equivalent implementations in state-of-the-art technologies in terms of energy consumption, delay, and area (the number of utilized devices). In addition, the thermal noise and variability effects are explained.

\subsection{Performance Evaluation}

The proposed FA is assessed and compared with the state-of-the-art CMOS \cite{7nmCMOS}, Magnetic Tunnel Junction MTJ \cite{MTJ1,MTJ2}, Spin Hall Effect SHE \cite{SHE2}, Domain Wall Motion DWM \cite{DWM}, and Spin-CMOS \cite{SPIN} based FA in terms of energy, delay, and  area (the number of utilized devices). In the evaluation and comparison, the following assumptions are made: (i) Excitation and detection cells are magnetoelectric (ME) cells. (ii) The ME's energy consumption and delay are \SI{14.4}{aJ} and \SI{0.42}{ns}, respectively \cite{Excitation_table_ref16}. (iv) SWs don't consume noticeable energy in the waveguide in comparison with the transducer energy consumption. (v) SWs are excited using pulse signals. Note that these assumptions might not reflect the reality of the spin wave technology because of its early stage development, and they might need to be re-evaluated in the future. 
\begin{table}[t]
\caption{FA Performance Comparison.}
\label{table:4}
\centering
  \begin{tabular}{|c|c|c|c|c|c|c|c|c|}
    \hline
     &  CMOS \cite{CMOS}   &  MTJ \cite{MTJ2} & SHE \cite{SHE2} & DWM \cite{DWM} & Spin-CMOS \cite{SPIN} & Coup. SW & Conv. SW & Prop. SW \tabularnewline \hline
     Energy (fJ) & $0.176$ & $5685$ & $4970$ & $74.8$ & $166.8$ & $0.129$ & $0.129$ & $0.1$  \tabularnewline
    \hline
     Delay (ns) & $0.1$ & $3.02$ &  $7$ & $0.88$ &  $3$ & $20.84$ & $2.86$ & $2.86$  \tabularnewline
    \hline
     Device No. &  $22$ &  $29$ & $26$ &  $68$&$34$&$9$&$9$&$7$
 \tabularnewline
    \hline
  \end{tabular}
\end{table}
The  SW FA delay is determined by adding the delay of $4$ ME cells because there are $4$ cells ($2$ excitation and $2$ detection cells) in the critical path to the SW propagation delay in the waveguide, which is extracted from micromagnetic simulation and it is \SI{1.18}{ns}. Therefore, the SW FA delay is \SI{2.86}{ns}. 

The straightforward approach to build a SW FA is by utilizing $3$ MAJ gates. However, as direct MAJ gate cascading  is not possible in the spin wave amplitude normalization is required, which can be performed either by converting SW gate outputs to charge domain and back  by means of two transducers or by directional couplers \cite{SW1}. As such we compare our implementation with both possible SW implementations, i.e.,  conversion based (Conv.) and coupler based (Coup.). Note that the  directional coupler  delay is \SI{20}{ns} \cite{SW1}. 

Table \ref{table:4} summarises the performance of the proposed SW FA and the considered contenders. As it can be observed from the Table, the SW FA saves $43$\% energy whereas it requires $28.6$x more delay when compared with the \SI{22}{nm} CMOS based FA design. Moreover, it consumes $4$ orders of magnitude less energy, and exhibits $5$\% and $59$\% less delay than the MTJ and SHE based FAs, respectively.  When compared with the DWM based FA it consumes $2$ orders of magnitude less energy at the expense of $3\times$ higher delay. Furthermore, the proposed FA consumes $3$ orders of magnitude less energy and exhibits $5$\% less delay in comparison with the Spin-CMOS based FA.  Last but not least, the proposed SW FA consumes $22.5$\% less energy than MAJ based SW implementations, while having the same and $10$x smaller delay than the Conv. and Coup. counterparts, respectively.  Note that the MTJ device number \cite{MTJ2} consists of $25$ transistor and $4$ MTJ, whereas the SHE device number \cite{MTJ2} consists of $23$ transistor and $3$ SHE-MTJ. Also, the DWM device \cite{MTJ2} consists of $20$ transistor, $4$ MTJ, and $2$ Domain Wall DW, whereas the SPIN-CMOS device \cite{MTJ2} consists of $28$ transistor, $4$ MTJ and $2$ DW. Note that the proposed FA needs the least number of devices, which indicates that it potentially requires a small chip real-estate. Note that we didn't consider the FA in \cite{logic1} in the comparison as up to date it has not been validated. Our attempts to do that by means MuMax3 failed as it relies on unattainable assumptions, e.g., output detection at the interference point, output initialization to $0$ before computing,  zero ME cell delay and  \SI{4.8}{aJ} power consumption.

\subsection{Variability and Thermal Effect}

Our main target in this paper is to validate the proof of concept of the proposed structure, regardless of variability and  thermal noise effects. However, in \cite{DC9,DC}, edge roughness and trapezoidal waveguide cross section were considered to test their effect on the gate functionality. It was demonstrated that the gate functions correctly under their presence and they only have a small effect \cite{DC9,DC}. Furthermore, the thermal noise effect was analyzed in \cite{DC} and it was concluded that noise has a negligible effect and the gate functions at different temperatures. Hence, we don't expect a noticeable effect of variability and thermal noise on the proposed structures. However, the investigation of such phenomena is subject of future work. 

\subsection*{Discussion}

The assessment indicates that the SW has the potential to advance the state-of-the-art in terms of energy as well as area consumption. However, there are still some open issues such as \cite{SW}:

\begin{itemize}

 \item Immature technology: MEs appear to be the right solution for SW excitation and detection as they have low power consumption potential and conceptually speaking can be utilized for both SW excitation and detection. However no actual ME experimental realization exists. 
 \item Cost and Complexity: Conceptually speaking SW devices can be scaled down to $nm$ as they must be greater or equal than the spin wave wavelength $\lambda$, thus properly behaving SWs  with wavelength in the $nm$ range are achievable. However, practical issues may need to be addressed in order to enable $nm$ range SW devices, including: Excitation and detection - $nm$ SWs cannot currently be generated and even if they would they cannot be distinguished from noise. 

\end{itemize} 
 We are confident however that if the other issues can be properly addressed SW based computation advantages potentially enabled the industry will find, as always, the way towards $nm$ range magnonic circuits and systems.

\section{Conclusions}
\label{sec:Conclusion}

A novel energy efficient spin wave based FA was proposed in this paper. The FA is implemented by making use of a Majority gate and $2$ XOR gates. In the proposed FA, two main detection mechanisms were utilized: phase detection for the Carry-out output detection and threshold detection for the Sum output detection. The correct functionality of the FA was validated by means of micromagnetic simulations and it was evaluated and compared with direct SW gate based implementation and five state-of-the-art technologies equivalent designs \SI{22}{nm} CMOS, MTJ, SHE, DWM and Spin-CMOS. It was demonstrated that the proposed FA consumes $22.5$\%, and $43$\% less energy than direct SW gate based implementations and \SI{22}{nm} CMOS, respectively and saves more than $3$ orders of magnitude in comparison with the state-of-the-art MTJ, SHE, DWM and Spin-CMOS based FA. Also, the proposed FA needs more than $22$\% less area in comparison with all designs.

\section*{Acknowledgement}

This work has received funding from the European Union's Horizon 2020 research and innovation program within the FET-OPEN project CHIRON under grant agreement No. 801055. It has also been partially supported by imec’s industrial affiliate program on beyond-CMOS logic. F.V. acknowledges financial support from the Research Foundation–-Flanders (FWO) through grant No.~1S05719N. 

\bibliography{Spin_Wave_Based_Full_Adder}

\end{document}